\documentclass[sigconf,natbib=true,screen=true,anonymous=false]{acmart}

\acmSubmissionID{482}

\usepackage{booktabs} 
\usepackage{amsmath}
\usepackage{graphics}
\usepackage{epsfig}
\usepackage{graphicx}
\usepackage[ruled,vlined, linesnumbered]{algorithm2e}
\usepackage{xcolor}
\usepackage[skip=0pt]{caption}
\usepackage{subfigure}
\usepackage{stfloats}
\usepackage{balance}
\usepackage{multirow}
\usepackage{mathrsfs}
\usepackage{acronym}
\usepackage{placeins}
\usepackage{tabularx}
\usepackage{makecell}
\usepackage{xcolor}
\usepackage[inline]{enumitem}
\usepackage{fancyhdr}
\usepackage{soul}

\SetKwInput{KwInput}{Input} 
\SetKwInput{KwOutput}{Output}
\usepackage{svg}
\usepackage{amsmath}
\AtBeginDocument{%
  \providecommand\BibTeX{{%
    \normalfont B\kern-0.5em{\scshape i\kern-0.25em b}\kern-0.8em\TeX}}}

\acrodef{KL}{Kullback-Leibler}
\acrodef{BCE}{Binary Cross Entropy}
\acrodef{BPR}{bayesian personalized ranking}
\acrodef{GCN}{graph convolutional network}
\acrodef{GNN}{graph neural network}
\acrodef{CL}{contrastive learning}
\acrodef{VAE}{variational autoencoder}
\acrodef{KL-divergence}{Kullback–Leibler divergence}

\allowdisplaybreaks

\newcommand{\headernodot}[1]{\vspace*{1mm}\noindent\textbf{#1}}
\newcommand{\header}[1]{\headernodot{#1.}}

\author{Xin Xin*}
\affiliation{%
  \institution{Shandong University}
  \city{Qingdao}
  \country{China}
}
\email{xinxin@sdu.edu.cn}

\author{Liu Yang*}
\affiliation{%
  \institution{Shandong University}
  \city{Qingdao}
  \country{China}
}
\email{yangliushirry@gmail.com}

\author{Ziqi Zhao}
\affiliation{%
  \institution{Shandong University}
  \city{Qingdao}
  \country{China}
}
\email{ziqizhao.work@gmail.com}

\author{Pengjie Ren}
\affiliation{%
  \institution{Shandong University}
  \city{Qingdao}
  \country{China}
}
\email{jay.ren@outlook.com}

\author{Zhumin Chen}
\affiliation{%
  \institution{Shandong University}
  \city{Qingdao}
  \country{China}
}
\email{chenzhumin@sdu.edu.cn}

\author{Jun Ma}
\affiliation{%
  \institution{Shandong University}
  \city{Qingdao}
  \country{China}
}
\email{majun@sdu.edu.cn}

\author{Zhaochun Ren$^{\dagger}$}
\email{z.ren@liacs.leidenuniv.nl}
\affiliation{%
  \institution{Leiden University}
  \city{Leiden}
  \country{Netherlands}
}

\makeatletter
\def\authornotetext#1{
\if@ACM@anonymous\else
    \g@addto@macro\@authornotes{
    \stepcounter{footnote}\footnotetext{#1}}
\fi}
\makeatother
\authornotetext{Equal contribution.}
\authornotetext{Corresponding author.}

\def\authors{Xin Xin, Liu Yang, Ziqi Zhao, Pengjie Ren, Zhumin Chen, Jun Ma, Zhaochun Ren}
\renewcommand{\shortauthors}{Xin Xin et al.}

\copyrightyear{2024}
\acmYear{2024}
\setcopyright{acmlicensed}
\acmConference[WSDM '24]{Proceedings of the 17th ACM International Conference on Web Search and Data Mining}{March 4--8, 2024}{Merida, Mexico}
\acmBooktitle{Proceedings of the 17th ACM International Conference on Web Search and Data Mining (WSDM '24), March 4--8, 2024, Merida, Mexico} 
\acmPrice{15.00}
\acmDOI{10.1145/3616855.3635823}
\acmISBN{979-8-4007-0371-3/24/03}

\theoremstyle{definition}
\newtheorem{definition}{Definition}[section]
\begin{document}
\begin{sloppypar}

\title[On the Effectiveness of Unlearning in Session-Based Recommendation]{On the Effectiveness of Unlearning in \\Session-Based Recommendation}    

\begin{abstract}
Session-based recommendation predicts users' future interests from previous interactions in a session. Despite the memorizing of historical samples, the request of unlearning, i.e., to remove the effect of certain training samples, also occurs for reasons such as user privacy or model fidelity. 
However, existing studies on unlearning are not tailored for the session-based recommendation. 
On the one hand, these approaches cannot achieve satisfying unlearning effects due to the collaborative correlations and sequential connections between the unlearning item and the remaining items in the session.
On the other hand, seldom work has conducted the research to verify the unlearning effectiveness in the session-based recommendation scenario.

In this paper, we propose SRU, a \textbf{s}ession-based \textbf{r}ecommendation \textbf{u}nlearning framework, which enables high unlearning efficiency, accurate recommendation performance, and improved unlearning effectiveness in session-based recommendation. 
Specifically, we first partition the training sessions into
separate sub-models according to the similarity across the sessions, then we utilize an attention-based aggregation
layer to fuse the hidden states according to the correlations between the session and the centroid of the data in the sub-model. To improve the unlearning effectiveness, we further propose three extra data deletion
strategies, including collaborative extra deletion (CED), neighbor extra deletion (NED), and random extra deletion (RED). Besides, we propose an evaluation metric that measures whether the unlearning sample can be inferred after the data deletion to verify the unlearning effectiveness. We implement SRU with three representative session-based recommendation models and conduct experiments on three benchmark datasets. Experimental results demonstrate the effectiveness of our methods. Codes and data
are available at \url{https://github.com/shirryliu/SRU-code}.
\end{abstract}

\ccsdesc[500]{Information systems~Recommender systems}
\ccsdesc[500]{Security and privacy~Web application security}

\keywords{Unlearning, Session-based Recommendation, Privacy Protection, Recommender System, Information Security}

\maketitle
\section{Introduction}
Session-based recommendation models have shown their effectiveness in predicting users’ future interests from memorized sequential interactions ~\cite{hu2018ecommerce,yuan2019music}. However, the ability to eliminate the influence of specific training samples, known as unlearning, also holds crucial significance.
From a legitimacy perspective, several data protection regulations have been proposed, such as the General Data Protection Regulation (GDPR)~\cite{mantelero2013eu} and the California Consumer Privacy Act (CCPA)~\cite{illman2019california}.
These legislative regulations emphasize individuals' right to have their private information removed from trained machine learning models.
As for user perspective, there has been a surge in research which proved that various user privacy information such as gender, age, and even political orientation could be inferred from historical interactions with a recommender system ~\cite{miars, DBLP:journals/corr/abs-2012-07805, GANLeaks}. Addressing privacy concerns, users might find it imperative to request the expunction of specific historical interactions.
Besides, a proficient recommendation model possesses the capacity to eliminate the impact of noisy training interactions to gain better performance.

\noindent\textbf{Machine unlearning.} 
Machine unlearning enables a model to forget certain data or patterns that it has previously learned. \emph{Exact unlearning} targets on completely eradicating the impact of the data to be forgotten as if they never occurred in the training process.
A straightforward exact unlearning method is to remove the targeted samples from the training dataset and then retrain the entire model from scratch. Unfortunately, this approach is hindered by its time-consuming and resource-intensive nature. To address this issue, existing methods ~\cite{bourtoule2021machine, brophy2021machine} focus on enhancing the efficiency of unlearning.  
One of the most representative unlearning methods is SISA~\cite{bourtoule2021machine}. The SISA initially divides the training dataset into disjoint shards of equal size. Subsequently, sub-models are trained on each shard independently. To formulate the final model prediction for a given data point, the predictions from every sub-model are aggregated through majority voting or average. In the event of an unlearning request, sorely the sub-model that was trained on the shard containing the unlearning data point is retrained, rather than the entire model. SISA achieves significant improvement in unlearning efficiency compared with the whole retraining.

\noindent \textbf{Challenges of unlearning in session-based recommendation.} 
\begin{figure}
 \centering
 \includegraphics[width=0.35\textwidth]{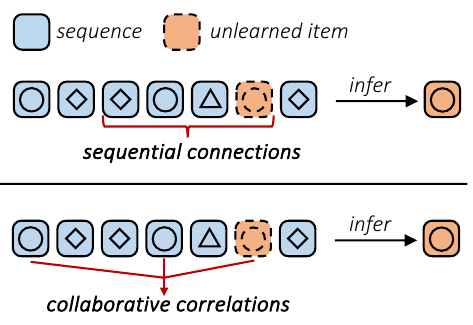}
 \caption{Exact unlearning is hard to achieve.  The unlearned item could still be inferred due to collaborative correlations and sequential connections across items in the session.}
 \label{fig:introduction}
\end{figure}
In the recommendation field, RecEraser~\cite{chen2022recommendation} applies the SISA framework to non-sequential collaborative filtering.
Nevertheless, we argue that there still exists the following key challenges for unlearning in session-based recommendation:

i). Exact unlearning is hard to achieve.
    Existing exact unlearning methods hold the assumption that the effect of unlearning samples would be completely removed if such samples do not exist in the retrained models. However, the assumption does not hold for session-based recommendation. Different from other domains such as image classification, where the correlations between training samples are sparse, there exists plenty of \emph{collaborative correlations} and \emph{sequential connections} across the interacted items in session-based recommendation. Consequently, simply removing the unlearning samples cannot achieve the exact unlearning effect, i.e., the unlearned item could still be inferred from the remaining items in the session, as shown in Figure \ref{fig:introduction}.
    
ii). Existing recommendation unlearning methods do not evaluate the unlearning effectiveness. 
    Existing methods ~\cite{chen2022recommendation, liu2022forgetting} mainly focus on the trade-off between recommendation performance and unlearning efficiency. However, seldom work conducted the evaluation regarding the unlearning effectiveness, i.e., to which extent the effect of unlearned samples is eliminated. The evaluation is especially important to verify the unlearning effectiveness of session-based recommendation, given the case that exact unlearning cannot be simply achieved.


\noindent \textbf{The proposed method.} In this paper, we propose \textbf{s}ession-based \textbf{r}ecommendation \textbf{u}nlearning (SRU), an unlearning framework tailored to session-based recommendation, achieving high unlearning efficiency, accurate recommendation, and improved unlearning effectiveness. 
Concretely, we first partition the training sessions into separate shards according to the similarity of the sessions and then a corresponding sub-model is trained upon each data shard. Such a data division strategy attempts to make similar sessions fall into the same shard, and thus each sub-model tends to learn a clustering of similar sequential patterns, resulting in improved recommendation performance. 
Utilizing the trained sub-models, we can obtain the hidden states which represent the session from the perspective of each sub-model. Then, an attention-based aggregation layer is trained to fuse the hidden states based on the correlations between the session and the centroid of the respective data shard. 

To address the first challenge, we propose three extra data deletion strategies, including collaborative extra deletion (CED), neighbor extra deletion (NED), and random extra deletion (RED), to further enhance the unlearning effectiveness. For the second challenge, we propose an evaluation metric that measures whether the unlearning sample can be inferred after data deletion. The intuition is that if the unlearning is highly effective, the unlearning sample should have a low probability of being inferred based on the remaining data. When an unlearning request occurs, e.g., a user may want to hide the click of a sensitive item, only the corresponding sub-model and the aggregation layer need to be retrained based on the deleted data to achieve efficient unlearning.

To verify the effectiveness of the proposed method, SRU is implemented on three representative session-based recommendation models including GRU4Rec~\cite{hidasi2015session}, SASRec~\cite{kang2018self}, and BERT4Rec~\cite{sun2019bert4rec}. We conduct a series of experiments on three benchmark datasets and the result shows the effectiveness of the proposed method.

\header{Contributions} To summarize, the main contributions lie in:
\begin{itemize}[leftmargin=*]
\item To the best of our knowledge, SRU is the first attempt that aims to address the machine unlearning problem for session-based recommendation. 
We propose three extra data deletion strategies to improve the unlearning effectiveness, and meanwhile, we use similarity-based clustering and attention-based aggregation to keep a high recommendation performance.
\item We propose an evaluation metric to verify the unlearning effectiveness of session-based recommendation. The key idea is that if the unlearning is effective, the unlearning sample should have a low probability of being inferred after data deletion. 
\item We conduct extensive experiments on three state-of-the-art session-based recommendation models and three benchmark datasets to show that SRU can achieve efficient and effective unlearning while keeping high recommendation performance.
\end{itemize}


\section{Related work}
In this section, we provide a literature review regarding session-based recommendation and machine unlearning.

\subsection{Session-Based Recommendation}
Session-based recommendation aims to capture a user’s dynamic interests from her/his past interactions in the session. Early Markov chains-based models ~\cite{7837843, FPMC, TBR, JMLR:v6:shani05a} predict a user's forthcoming interests according to the last interaction in the given session. More recently, deep neural network models have been utilized to capture complex sequential signals to improve session-based recommendation. The representative session-based recommendation models can be categorized into recurrent neural network (RNN)-based models ~\cite{Sequential-User-Based-Recurrent,Hidasi_2018}, convolutional neural network (CNN)-based models ~\cite{tang2018personalized}, attention-based models ~\cite{kang2018self,sun2019bert4rec}, and graph-based models ~\cite{Wu_Tang_Zhu_Wang_Xie_Tan_2019}. Besides, self-supervised learning ~\cite{self-supervised} and contrastive learning ~\cite{Contrastive,liu2021contrastive} have also been applied to improve session-based recommendation and plenty of models have been emerged.

In this paper, we propose a framework that enables effective and efficient unlearning for various session-based recommendation models, other than developing a new specific model. We adopt three representative models, GRU4Rec, SASRec and BERT4Rec, as backbone models for the experiments.
\subsection{Machine Unlearning}
The concept of machine unlearning was first proposed by~\cite{cao2015towards}, in response to the requirement of ``the right to be forgotten''. Unlearning methods can be broadly categorized into approximate unlearning methods and exact unlearning methods.

Approximate unlearning ensures that the performance of the unlearned model closely aligns with that of a retrained model. This reduces the time and computational cost of unlearning, but at the potential expense of weaker privacy assurances.
The approximation can be achieved through differential privacy techniques, such as certified unlearning ~\cite{MUsurvey}. 
For instance, ~\cite{ullah2021machine} introduced an unlearning method based on noisy stochastic gradient descent, whereas ~\cite{pmlr-v119-guo20c} achieved certified unlearning based on Newton updates. 
~\cite{cao2015towards} proposed to use the gradient surgery which updates the model parameters using the negative gradient of the unlearning samples. ~\cite{pmlr-v130-izzo21a} utilized a probabilistic model to approximate the unlearning process.
~\cite{mazhuo, chen2021lightweight} proposed to perturb the gradients or model weights through the inverse Hessian matrix, which may incur additional computational overheads.

Exact unlearning attempts to completely remove the effect of the unlearning samples as if they have never occurred in the training process, providing a stronger privacy guarantee. However, such methods could require the model to be retrained from scratch, which is computationally expensive and time-consuming. The most representative method for efficient exact unlearning is SISA ~\cite{bourtoule2021machine} since only the sub-model trained on the corresponding data shard would be retrained for an unlearning request. ~\cite{Chen_2022} adapted SISA for unlearning in graph neural networks. 
~\cite{gupta2021adaptive} modified the SISA algorithm to work for sequences of deletion requests.
Another kind of method for exact unlearning involves selective influence estimators ~\cite{gif}, which calculate the influence of the unlearning samples on the model parameters. Although such influence-based methods are effective in terms of privacy preservation, the high computational cost limits their application for real-world scenarios \cite{MUsurvey}. 



Recently, unlearning in the recommendation scenario tends to attract more research attention. Unlearning can not only help to protect user privacy but also improve recommendation models through eliminating the effect of noisy data and misleading information ~\cite{shaik2023exploring}.
~\cite{liu2022forgetting} and ~\cite{xu2023netflix} proposed to use fine-tuning and the alternative least square algorithm for unlearning acceleration. ~\cite{chen2022recommendation} and ~\cite{li2022making} extended the ideas of the SISA algorithm for collaborative filtering. However, none of the existing methods is tailored for session-based recommendation. Besides, existing methods mainly focus on the unlearning efficiency, while failing to verify the effectiveness of the unlearning, i.e., to which extent the effect of the unlearning sample is removed. 


\section{Task formulation}
In this section, we first formulate the task of session-based recommendation, upon which we define the task of item-level unlearning in a session. Then we identify the unlearning challenges. 
\subsection{Notations and Definitions}
Session-based recommendation aims to predict the user's potential next action given previous interacted items in the session. We formulate the task as follows:
\begin{definition}[Session-based recommendation]
Let $\mathcal{V}=\{v_{1},v_{2},...,v_{|\mathcal{V}|}\}$ be the set of items, $\mathcal{D}$ denotes the training interaction sessions.  $\mathcal{S}_{i}=[v^{i}_{1},...,v^{i}_{t},...,v^{i}_{n}] \in \mathcal{D}$  denotes the $i$-th specific interaction session in $\mathcal{D}$, where $v^{i}_t \in \mathcal{V}$ is the item interacted by the user at time step $t$, and $n$ is the current length of the session. 
Given the historical sequence $\mathcal{S}_{i}$, the interaction probability over candidate item $v$ at time step $n+1$ can be formalized as:
\begin{equation}
p^i_v=p(v^{i}_{n+1}=v|\mathcal{S}_{i},\mathcal{D})=\mathcal{M}(v|\mathcal{S}_i,\mathcal{D}),
\end{equation}
where $\mathcal{M}$ denotes the involved recommendation model, e.g., GRU4Rec \cite{Hidasi_2018} and SASRec \cite{kang2018self}. At the prediction stage, session-based recommenders select the items with the highest top-$K$ probability $p^i_v$ as the recommendation list for the user.

\end{definition}

For privacy considerations or recommendation utility, an unlearning request could occur to remove the effect of certain training samples. As an illustration, a user may want to revoke some misclicks in an interaction session since the misclicks can downgrade the recommendation quality or a user could also request to hide the click of certain sensitive items for private concerns. 
In this paper, we focus on the item-level unlearning in session-based recommendation, which is defined as follows:

\begin{definition}[Item-level unlearning]
We denote $v^{i}_{j} \in \mathcal{S}_{i}$ to be the unlearning item that the user wants to revoke in the session $\mathcal{S}_{i}$.
The goal of item-level unlearning is to obtain an unlearned model $\mathcal{M}_{u}$. Ideally, the unlearning sample $v^{i}_{j}$ should have no effect on the unlearned model $\mathcal{M}_{u}$ as if $v^{i}_{j}$ never occurred in the session.
\end{definition}
Note that besides the item-level unlearning, there could also be the request for session-level unlearning, i.e., to remove the effect of a whole interaction session. In this paper, we focus on item-level unlearning while the proposed framework can also support session-level unlearning. We leave the further investigation of session-level unlearning as one of our future directions.



\subsection{Challenges}
\subsubsection{Exact unlearning is hard to achieve}
\label{subsub:exact is hard}
Existing exact unlearning methods are mainly applied in fields such as image classification ~\cite{bourtoule2021machine}, where data points are relatively independent of each other. 
In this circumstance, existing unlearning methods hold a belief that the effect of unlearning samples would be perfectly removed if the samples do not exist in the retrained models. 
However, due to the fact that there exists plenty of \emph{collaborative correlations} and \emph{sequential connections} in item interactions of session-based recommendation, simply removing the unlearning sample cannot achieve the expected effect that the unlearning sample has never occurred in the training data. For example, a user may want to hide the click of a sensitive item in a session while simply removing the unlearning item in the session cannot achieve a satisfying effect since the sensitive item could still be inferred from the deleted data due to the existence of collaborative correlations or sequential corrections. The challenge can be formalized as:
\begin{equation}
   \text{exact unlearning}\neq\mathcal{M}_u(v|\mathcal{S}_i',\mathcal{D}\backslash\mathcal{S}_i\cup\mathcal{S}_{i}'), \text{where } \mathcal{S}_i'=\mathcal{S}_i\backslash v_j^i.
\end{equation}



\subsubsection{Unlearning effectiveness is not well defined}
Existing recommendation unlearning methods ~\cite{chen2022recommendation, liu2022forgetting} mainly investigated the trade-off between recommendation performance and unlearning efficiency. 
As pointed out in section \ref{subsub:exact is hard}, exact unlearning is hard to achieve in session-based recommendation. 
Under this situation, the evaluation of unlearning effectiveness, i.e., to which extent the influence of unlearning samples is eliminated, assumes particular significance.
However, seldom work has conducted the evaluation regarding the unlearning effectiveness in the field of session-based recommendation. 

Despite the above two specific challenges of unlearning in session-based recommendation, the model performance (i.e., recommendation accuracy) and unlearning efficiency are also key factors that need to be optimized.



\section{Methodology}
In this section, we describe the detail of the proposed SRU framework. As shown in Figure \ref{fig:model overview}, SRU is composed of \emph{session partition}, \emph{attentive aggregation}, and \emph{data deletion}. The session partition module aims to divide the training sessions into disjoint data shards and then sub-models are trained on each shard. Based on the hidden states coming from different sub-models, the attentive aggregation module fuses the hidden states for the final prediction. The data deletion module aims to improve the unlearning effectiveness.   
When an item-level unlearning request comes, the data deletion module first applies extra data deletion strategies to the corresponding session. Then only the sub-model and the aggregation module are retrained, achieving efficient unlearning.    




\subsection{Session Partition}
\label{subsec:session partition}
One keystone to generating the next-item recommendation is learning signals from similar sessions. To this end, session similarity is important for recommendation accuracy. 
Consequently, in the session partition module, similar sessions are expected to be divided into the same data shard and thus can be trained in one sub-model. Such a division strategy can help to improve the recommendation performance since it enables more knowledge transfer within each shard.  


To achieve the described division strategy, an additional session-based recommendation model $\mathcal{M}_p$ (e.g., GRU4Rec\cite{hidasi2015session}) is pre-trained on $\mathcal{D}$ to obtain all training sessions' hidden states firstly. Then a $k$-means clustering method based on the pre-trained hidden states is used to divide training sessions. 
More specifically, the input of the session partition module includes the pre-trained hidden states, the number of partition shards $\mathcal{K}$, and the maximum number of sessions in each shard $\delta$. 
The distance between session pairs is defined as the Euclidean distance of their hidden states. 
$\mathcal{K}$ sessions are randomly selected as centroids at first, then distances between sessions and centroids are calculated. Subsequently, the sessions are assigned to the shard sequentially according to the ascending order of distances.
If one shard is unavailable (i.e., the number of sessions within the shard is larger than $\delta$),  the next session is assigned to the nearest available shard. After that, the new centroids are calculated as the mean of all sessions' hidden states in each corresponding shard. 
The above process is repeated until the centroids are no longer updated. 
Then we obtain the balanced-partition session as $\bigcap_{k \in [\mathcal{K}]}\mathcal{D}_{k}=\varnothing$ and $\bigcup_{k \in [\mathcal{K}]}\mathcal{D}_{k}=\mathcal{D}$.
Then sub-models are trained on the data shard separately.





\subsection{Attentive Aggregation}
\label{subsec:aggregation}
\begin{figure}
 \centering
 \includegraphics[width=0.5\textwidth]{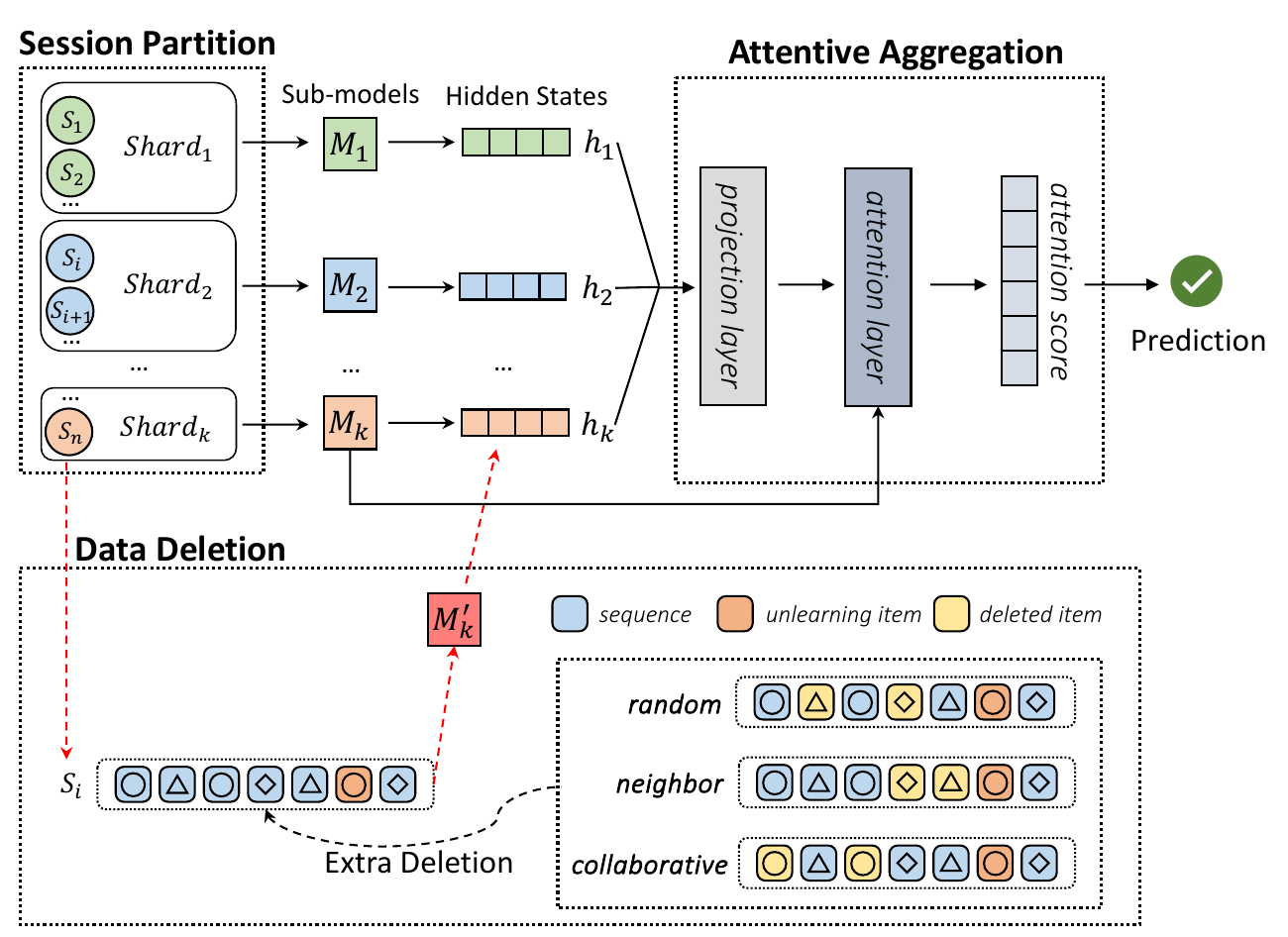}
 \caption{Overview of the proposed SRU framework. SRU is composed of session partition, attentive aggregation and data deletion modules.}
 \label{fig:model overview}
\end{figure}
Based on the session partition, each sub-model tends to learn a clustering of similar sequential patterns. The attentive aggregation module aims to fuse the hidden states from each sub-model for the final prediction, which consists of a projection layer, an attention layer, and an output layer.

\subsubsection{Projection layer} Given a session we compute its hidden representation $\mathbf{h}_{k}\in\mathbb{R}^{d}$ using each sub-model $\mathcal{M}_{k}$ trained on $\mathcal{D}_{k}$. 
Since sub-models are trained separately,  the hidden representations could embed in different vector spaces. In order to utilize the knowledge of every sub-model, we need to project the hidden representations into a common space. 
Specifically, a linear transfer layer is used to conduct the projection:
\begin{equation}
\mathbf{h}'_{k} = \mathbf{W}_{k}\mathbf{h}_{k} + \mathbf{b}_{k},
\end{equation}
where $\mathbf{W}_k\in\mathbb{R}^{d\times d}$ and $\mathbf{b}_{k}\in\mathbb{R}^d$ are projection parameters. Note that each sub-model $\mathcal{M}_k$ has a corresponding $\mathbf{W}_k$ and $\mathbf{b}_k$.

Besides, the data centroid of $\mathcal{D}_k$ is also projected as
\begin{equation}
\mathbf{c}'_{k} = \mathbf{W}_{k}\mathbf{c}_{k} + \mathbf{b}_{k},
\end{equation}
where $\mathbf{c}_k$ denotes the original centroid representation computed from $\mathbf{h}_k$.
The data centroid representation is used in the following attention layer.  
\subsubsection{Attention layer}
The attention layer aims to compute the importance of each sub-model for a given session. 
\cite{chen2022recommendation} also used an attention layer to fuse user and item embeddings for unlearning in collaborative filtering. However, their method cannot be applied to session-based recommendation since their attention is solely based on either user or item embedding.
While in the session-based recommendation, we argue that the attention should be based on the correlations between the session and the data centroid (i.e., the attention layer should have two input sources corresponding to the session representation and the centroid representation, as shown in Figure \ref{fig:model overview}). 

To this end, we define the attention score for sub-model $\mathcal{M}_k$ as
\begin{equation}
a_{k} = softmax(\mathbf{g}\cdot ReLU(\mathbf{W}'\odot(\mathbf{h}'_k\odot \mathbf{c}'_k) + \mathbf{b}')),
\end{equation}
where $\mathbf{W}'\in \mathbb{R}^{d\times f}$, $\mathbf{b}' \in \mathbb{R}^{f}$ and $\mathbf{g} \in \mathbb{R}^{f}$ are learnable attention parameters. $f$ is the size of the attention layer. $\odot$ denotes element-wise product and $\cdot$ denotes the inner product.

Based on the attention score, the final representation of a session is formulated as  
\begin{equation}
\mathbf{h}^{f} = \sum_{k=1}^\mathcal{K} a_{k} \mathbf{h}'_{k}.
\end{equation}



\subsubsection{Output layer} Based on the final aggregated hidden representation $\mathbf{h}^{f}$, a two-layer feed-forward network with ReLU activation is used to produce the output distribution over candidate items. 
The attentive aggregation module is trained with the cross-entropy loss over the output distribution. 

\subsection{Data Deletion}
\label{subsec:deletion}


The data deletion module aims to improve the unlearning effectiveness. 
For an item-level unlearning request, conventional unlearning methods just remove the unlearning sample, while it is still possible that the removed sample can be inferred again from the remaining interactions in the session due to the existence of sequential connections and collaborative correlations. 
To address the problem, we propose three strategies, namely Collaborative Extra Deletion(CED), Neighbor Extra Deletion(NED) and Random Extra Deletion(RED).  

From the view of collaborative correlations, we propose CED which deletes extra 
items based on the similarities between the unlearning item and other items in the session. Given the target unlearning item $v_{j}^i$ in session $\mathcal{S}_i$, the item similarity is calculated according to the Euclidean distance between item embeddings obtained from the pre-trained model $\mathcal{M}_p$. After that, the items in the session are sorted by the  ascending order of the distances and the most $\mathcal{N}$ similar items are also removed from the session. Finally, the corresponding sub-model and the aggregation module are re-trained. The unlearned model can be formalized as 
\begin{equation}
\mathcal{M}_u(v|\mathcal{S}_i'',\mathcal{D}\backslash\mathcal{S}_i\cup\mathcal{S}_{i}''), \text{where } \mathcal{S}_i''=\mathcal{S}_i\backslash \text{CED}(v_j^i).
\end{equation}





As for sequential connections, NED is proposed to remove the $\mathcal{N}$ nearest items in front of the unlearning item in chronological order. While in RED, we randomly choose $\mathcal{N}$ extra items to delete within the session.








\subsection{Unlearning Effectiveness Evaluation}
\label{subsec:evaluation}
Item-level unlearning is a common request, for example, a user may want to hide the click of a sensitive item in a session or may dislike an item anymore.
To this end, if the unlearning is effective, the unlearned item should not be inferred from the remaining items in the session or the item should not be recommended to the user again in the near future. 

To this end, we define one unlearning effectiveness evaluation metric as the hit ratio (i.e., HIT@$K$) which measures whether the unlearned item would occur in the top-$K$ recommendation list based on the remaining interactions in the session using the unlearned model $\mathcal{M}_u$.
Such the evaluation metric can also be seen as the performance of a membership inference attack \cite{shokri2017membership} which attempts to infer the unlearning items from the remaining data. If HIT@$K$ is high, it means the unlearned item has a high probability of being re-recommended or being inferred again. On the contrary, a lower HIT@$K$ implies the unlearned model $\mathcal{M}_u$ has unlearned the item well enough and achieves a better unlearning effectiveness.

\begin{table}[t]
\caption{Statistics of the datasets (after preprocessing).}
\label{table:data}
\centering
    \begin{tabular}{l rrr}
    \toprule
    Dataset & \#users  & \#items & \#actions \\ 
    \midrule
    Amazon Beauty  & 52,024  & 57,289  & 0.4M \\
    Amazon Games  & 31,013  & 23,715 & 0.3M \\
    Steam & 334,730  & 13,047  & 3.7M \\
    \bottomrule
    \end{tabular}
\end{table}
\section{Experiments}

In this section, we conduct experiments on three benchmark datasets to verify the effectiveness of SRU. We aim to answer the following research questions:

\textbf{RQ1:} How is the recommendation performance of SRU when instantiated with different session-based recommendation models?

\textbf{RQ2:} How is the unlearning effectiveness of SRU?

\textbf{RQ3:} How is the unlearning efficiency of SRU?

More ablation experiments about how do designs of SRU affect the performance can be seen in Appendix. 
\subsection{Experimental Settings}

\subsubsection{Datasets}
Experiments are conducted on three publicly accessible datasets: \emph{Amazon Beauty}, \emph{Games} and \emph{Steam}. The two Amazon datasets\footnote{\url{https://jmcauley.ucsd.edu/data/amazon/}} are a series of product review datasets crawled from Amazon.com. In this work, we consider two item categories including ``Beauty'' and ``Games''. The Steam dataset\footnote{\url{https://steam.internet.byu.edu/}} is collected from a large online video game distribution platform. For all datasets, we follow the same data prepossessing as \cite{kang2018self}.
Table \ref{table:data} shows the statistics of the datasets.


\subsubsection{Recommendation performance evaluation}
We adopt cross-validation to evaluate the performance of the proposed methods. The ratio of training, validation, and test set is 8:1:1. We randomly sample 80\% of the sessions as the training set. For validation and test sets, The evaluation is done for validation and test sets by providing the interactions in a session one by one and checking the rank of the next ground-truth item. The ranking is performed among the whole item set.

To evaluate recommendation performance, we adopt two common top-$K$ metrics: Recall@$K$ and NDCG@$K$. Recall@$K$ measures whether the ground-truth item is in the top-$K$ positions of the recommendation list~\cite{xin2020self}. NDCG@$K$ is a weighted metric that assigns higher scores to top-ranked positions ~\cite{jarvelin2002cumulated}. We use the metric HIT described in section \ref{subsec:evaluation} to evaluate unlearning effectiveness.

\subsubsection{Baselines}
\begin{table*}[t]
\caption{Recommendation performance comparison after unlearning 10\% of data in each shard. The extra deletion number $\mathcal{N}$ ranges from 1 to 5. Best results other than Retrain are highlighted in bold. ``N'' is short for NDCG and ``R'' is short for Recall.}
\label{table:performance10}
\centering
    \begin{tabular}{ccccc ccccc ccccc cc}
    
    \toprule
    \multicolumn{2}{c}{\multirow{2}*{\textbf{Beauty}}}& \multicolumn{4}{c}{\textbf{GRU4Rec}}& \multicolumn{4}{c}{\textbf{SASRec}}& \multicolumn{4}{c}{\textbf{BERT4Rec}}\\
    \cmidrule(lr){3-6}\cmidrule(lr){7-10}\cmidrule(lr){11-14}
    \multicolumn{2}{c}{}&\textbf{N@10}&\textbf{N@20}&\textbf{R@10}&\textbf{R@20}&\textbf{N@10}&\textbf{N@20}&\textbf{R@10}&\textbf{R@20}&\textbf{N@10}&\textbf{N@20}&\textbf{R@10}&\textbf{R@20}\\
    \midrule
    \multicolumn{2}{c}{\textbf{Retrain}} & 0.0327 & 0.0382 & 0.0550 & 0.0768 & 0.0399 & 0.0450 & 0.0632 & 0.0835 & 0.0314 & 0.0380 & 0.0558 & 0.0816\\
    \multicolumn{2}{c}{\textbf{SISA}} & 0.0289 & 0.0328 & 0.0460 & 0.0615 & 0.0271 & 0.0307 & 0.0428 & 0.0571 & 0.0259 & 0.0310 & 0.0464 & 0.0666\\
    \multicolumn{2}{c}{\textbf{SRU-R}} & 0.0304 & \textbf{0.0347} & 0.0489 & 0.0662 & 0.0280 & \textbf{0.0323} & 0.0448 & \textbf{0.0617} & 0.0292 & 0.0341 & 0.0509 & 0.0704 \\
    \multicolumn{2}{c}{\textbf{SRU-C}} & 0.0286 & 0.0330 & 0.0468 & 0.0643 & \textbf{0.0280} & 0.0320 & \textbf{0.0456} & 0.0616 & \textbf{0.0293} & \textbf{0.0348} & \textbf{0.0525} & \textbf{0.0743} \\
    \multicolumn{2}{c}{\textbf{SRU-N}} & \textbf{0.0306} & 0.0346 & \textbf{0.0506} & \textbf{0.0668} & 0.0274 & 0.0312 & 0.0440 & 0.0591 & 0.0291 & 0.0346 & 0.0507 & 0.0726 \\
    \Xcline{1-1}{0.4pt}
    \Xhline{0.6pt}

    \toprule
    \multicolumn{2}{c}{\multirow{2}*{\textbf{Steam}}}& \multicolumn{4}{c}{\textbf{GRU4Rec}}& \multicolumn{4}{c}{\textbf{SASRec}}& \multicolumn{4}{c}{\textbf{BERT4Rec}}\\
    \cmidrule(lr){3-6}\cmidrule(lr){7-10}\cmidrule(lr){11-14}
    \multicolumn{2}{c}{}&\textbf{N@10}&\textbf{N@20}&\textbf{R@10}&\textbf{R@20}&\textbf{N@10}&\textbf{N@20}&\textbf{R@10}&\textbf{R@20}&\textbf{N@10}&\textbf{N@20}&\textbf{R@10}&\textbf{R@20}\\
    \midrule
    \multicolumn{2}{c}{\textbf{Retrain}} & 0.0495 & 0.0631 & 0.0947 & 0.1489 & 0.0539 & 0.0679 & 0.1016 & 0.1574 & 0.0593 & 0.0742 & 0.1116 & 0.1711\\
    \multicolumn{2}{c}{\textbf{SISA}} & 0.0471 & 0.0601 & 0.0898 & 0.1412 & 0.0457 & 0.0581 & 0.0863 & 0.1357 & 0.0482 & 0.0615 & 0.0932 & 0.1460\\
    \multicolumn{2}{c}{\textbf{SRU-R}} & \textbf{0.0490} & \textbf{0.0621} & \textbf{0.0924} & 0.1444 & \textbf{0.0485} & \textbf{0.0614} & \textbf{0.0914} & \textbf{0.1431} & \textbf{0.0577} & \textbf{0.0722} & \textbf{0.1077} & \textbf{0.1652}\\
    \multicolumn{2}{c}{\textbf{SRU-C}} & 0.0484 & 0.0616 & 0.0916 & \textbf{0.1445} & 0.0476 & 0.0604 & 0.0901 & 0.1411 & 0.0576 & 0.0720 & 0.1075& 0.1648\\
    \multicolumn{2}{c}{\textbf{SRU-N}} & 0.0480 & 0.0612 & 0.0916 & 0.1442 & 0.0480 & 0.0608 & 0.0906 & 0.1414 & 0.0567 & 0.0710 & 0.1067 & 0.1636\\
    \Xcline{1-1}{0.4pt}
    \Xhline{0.6pt}

    \toprule
    \multicolumn{2}{c}{\multirow{2}*{\textbf{Games}}}& \multicolumn{4}{c}{\textbf{GRU4Rec}}& \multicolumn{4}{c}{\textbf{SASRec}}& \multicolumn{4}{c}{\textbf{BERT4Rec}}\\
    \cmidrule(lr){3-6}\cmidrule(lr){7-10}\cmidrule(lr){11-14}
    \multicolumn{2}{c}{}&\textbf{N@10}&\textbf{N@20}&\textbf{R@10}&\textbf{R@20}&\textbf{N@10}&\textbf{N@20}&\textbf{R@10}&\textbf{R@20}&\textbf{N@10}&\textbf{N@20}&\textbf{R@10}&\textbf{R@20}\\
    \midrule
    \multicolumn{2}{c}{\textbf{Retrain}} & 0.0401 & 0.0495 & 0.0747 & 0.1122 & 0.0479 & 0.0580 & 0.0864 & 0.1268 & 0.0474 & 0.0596 & 0.0921 & 0.1406\\
    \multicolumn{2}{c}{\textbf{SISA}} & 0.0324 & 0.0377 & 0.0564 & 0.0776 & 0.0267 & 0.0318 & 0.0459 & 0.0661 & 0.0322 & 0.0402 & 0.0629 & 0.0948 \\
    \multicolumn{2}{c}{\textbf{SRU-R}} & \textbf{0.0357} & 0.0424 & \textbf{0.0621} & 0.0887 & \textbf{0.0333} & \textbf{0.0405} & \textbf{0.0596} & \textbf{0.0883} & \textbf{0.0395} & \textbf{0.0497} & \textbf{0.0752} & \textbf{0.1159} \\
    \multicolumn{2}{c}{\textbf{SRU-C}} & 0.0342 & 0.0410 & 0.0614 & 0.0887 & 0.0314 & 0.0378 & 0.0570 & 0.0824 & 0.0363 & 0.0462 & 0.0690 & 0.1084 \\
    \multicolumn{2}{c}{\textbf{SRU-N}} & 0.0352 & \textbf{0.0424} & 0.0620 & \textbf{0.0909} & 0.0321 & 0.0393 & 0.0566 & 0.0851 & 0.0384 & 0.0488 & 0.0730 & 0.1146 \\
    
    \bottomrule
    \end{tabular}
\end{table*}
SRU is implemented with three representative session-based recommendation models: GRU4Rec \cite{hidasi2015session}, SASRec \cite{kang2018self} and BERT4Rec \cite{sun2019bert4rec}.
\begin{itemize}[leftmargin=*,nosep]
\item \textbf{GRU4Rec} \cite{hidasi2015session}: This method utilizes gated recurrent units (GRU) to model user interaction sequences.
\item \textbf{SASRec} \cite{kang2018self}: This model is attention-based and uses the Transformer \cite{vaswani2017attention} decoder for session-based recommendation.

\item \textbf{BERT4Rec} \cite{sun2019bert4rec}: This model employs deep bidirectional self-attention to model interaction sequences.

\end{itemize}
To enable unlearning, every model is trained with:
\begin{itemize}[leftmargin=*,nosep]
\item \textbf{Retrain:} This method retrains the whole model from scratch on the remaining dataset. It's computationally expensive.

\item \textbf{SISA:} This is a fundamental exact unlearning method that randomly splits the data and averages the outputs of the sub-model. 


\item \textbf{SRU-N:} This is SRU with neighbor extra deletion (NED).

\item \textbf{SRU-R:} This is SRU with random extra deletion (RED).

\item \textbf{SRU-C:} This is SRU with collaborative extra deletion (CED).
\end{itemize}
Note that we do not compare with RecEraser \cite{chen2022recommendation} since it is proposed for non-sequential collaborative filtering and their data partition methods cannot be applied for session-based recommendation since the session-based recommender does not explicitly model user identifiers.

\subsubsection{Hyperparameter settings}
The model input is the last 10 interacted items for Beauty, and the last 20 interacted items for Games and Steam. 
We pad the sequences with a padding token for shorter sessions. 
The Adam optimizer~\cite{kingma2014adam} is used to train all models, with batches of size 256. The learning rate for the aggregation layer is tuned among [1e-3, 1e-2]. The default number of data shard is set as $\mathcal{K}=8$. The extra data deletion number for unlearning is ranged from 1 to 5. The other hyperparameters are set as the recommended settings of their original papers.

\subsection{Recommendation Performance (RQ1)}
Table \ref{table:performance10} shows the top-$K$ recommendation performance 
of different unlearning methods when 10$\%$ random sessions need items to be unlearned in each shard. We can see that the proposed SRU always performs better than SISA even though SRU has removed more training data. This is because when training data is unlearned, the performance of all sub-models are degraded for the training data is smaller, while SRU groups similar sessions in a shard which makes the model share more collaborative information to gain better recommendation performance. And it makes sense that Retrain always gets the highest scores for it can retrain the whole model on all remaining data, but sacrifices efficiency.

Besides, we also conduct experiments to see the recommendation performance when there is no unlearning request. Table \ref{table:performance} shows the recommendation performance with full training data. The Retrain model achieves the best scores. Compared with SISA, we can see that the proposed SRU always gains much better recommendation performance. For instance, on the BERT4Rec model trained on the Steam dataset, the NDCG@10 for SRU is 0.0575, while the corresponding result is 0.0492 for SISA, achieving a 16.9\% improvement. The observation demonstrates the effectiveness of the session partition and the attentive aggregation of SRU.  

To conclude, the proposed SRU achieves better recommendation performance than baseline methods on both unlearning request scenarios and full data scenarios.



\begin{table*}[t]
\caption{Recommendation performance comparison without unlearning request.
SRU denotes the proposed SRU framework without extra data deletion. 
Best results other than Retrain are highlighted in bold. 
``N'' is short for NDCG and ``R'' is short for Recall.}
\label{table:performance}
\centering
    \begin{tabular}{ccccc ccccc ccccc cc}
    
    \toprule
    \multicolumn{2}{c}{\multirow{2}*{\textbf{Beauty}}}& \multicolumn{4}{c}{\textbf{GRU4Rec}}& \multicolumn{4}{c}{\textbf{SASRec}}& \multicolumn{4}{c}{\textbf{BERT4Rec}}\\
    \cmidrule(lr){3-6}\cmidrule(lr){7-10}\cmidrule(lr){11-14}
    \multicolumn{2}{c}{}&\textbf{N@10}&\textbf{N@20}&\textbf{R@10}&\textbf{R@20}&\textbf{N@10}&\textbf{N@20}&\textbf{R@10}&\textbf{R@20}&\textbf{N@10}&\textbf{N@20}&\textbf{R@10}&\textbf{R@20}\\
    \midrule
    \multicolumn{2}{c}{\textbf{Retrain}} & 0.0340 & 0.0396 & 0.0580 & 0.0801 & 0.0419 & 0.0471 & 0.0668 & 0.0871 & 0.0366 & 0.0430 & 0.0634 & 0.0889\\
    \multicolumn{2}{c}{\textbf{SISA}} & 0.0309 & 0.0347 & 0.0490 & 0.0642 & 0.0283 & 0.0323 & 0.0453 & 0.0614 & 0.0271 & 0.0323 & 0.0491 & 0.0694\\
    \multicolumn{2}{c}{\textbf{SRU}} & \textbf{0.0313} & \textbf{0.0360} & \textbf{0.0507} & \textbf{0.0691} & \textbf{0.0296} & \textbf{0.0341} & \textbf{0.0477} & \textbf{0.0655} & \textbf{0.0316} & \textbf{0.0371} & \textbf{0.0558} & \textbf{0.0777} \\
    \Xcline{1-1}{0.4pt}
    \Xhline{0.6pt}

    \toprule
    \multicolumn{2}{c}{\multirow{2}*{\textbf{Steam}}}& \multicolumn{4}{c}{\textbf{GRU4Rec}}& \multicolumn{4}{c}{\textbf{SASRec}}& \multicolumn{4}{c}{\textbf{BERT4Rec}}\\
    \cmidrule(lr){3-6}\cmidrule(lr){7-10}\cmidrule(lr){11-14}
    \multicolumn{2}{c}{}&\textbf{N@10}&\textbf{N@20}&\textbf{R@10}&\textbf{R@20}&\textbf{N@10}&\textbf{N@20}&\textbf{R@10}&\textbf{R@20}&\textbf{N@10}&\textbf{N@20}&\textbf{R@10}&\textbf{R@20}\\
    \midrule
    \multicolumn{2}{c}{\textbf{Retrain}} & 0.0490 & 0.0623 & 0.0943 & 0.1475 & 0.0541 & 0.0682 & 0.1021 & 0.1584 & 0.0613 & 0.0764 & 0.1142 & 0.1742\\
    \multicolumn{2}{c}{\textbf{SISA}} & 0.0473 & 0.0600 & 0.0903 & 0.1412 & 0.0463 & 0.0586 & 0.0874 & 0.1366 & 0.0492 & 0.0628 & 0.0947 & 0.1489\\
    \multicolumn{2}{c}{\textbf{SRU}} & \textbf{0.0474} & \textbf{0.0603} & \textbf{0.0903} & \textbf{0.1415} & \textbf{0.0483} & \textbf{0.0615} & \textbf{0.0915} & \textbf{0.1437} & \textbf{0.0575} & \textbf{0.0718} & \textbf{0.1072} & \textbf{0.1643}\\
    \Xcline{1-1}{0.4pt}
    \Xhline{0.6pt}

    \toprule
    \multicolumn{2}{c}{\multirow{2}*{\textbf{Games}}}& \multicolumn{4}{c}{\textbf{GRU4Rec}}& \multicolumn{4}{c}{\textbf{SASRec}}& \multicolumn{4}{c}{\textbf{BERT4Rec}}\\
    \cmidrule(lr){3-6}\cmidrule(lr){7-10}\cmidrule(lr){11-14}
    \multicolumn{2}{c}{}&\textbf{N@10}&\textbf{N@20}&\textbf{R@10}&\textbf{R@20}&\textbf{N@10}&\textbf{N@20}&\textbf{R@10}&\textbf{R@20}&\textbf{N@10}&\textbf{N@20}&\textbf{R@10}&\textbf{R@20}\\
    \midrule
    \multicolumn{2}{c}{\textbf{Retrain}} & 0.0416 & 0.0516 & 0.0778 & 0.1174 & 0.0493 & 0.0604 & 0.0877 & 0.1319 & 0.0471 & 0.0595 & 0.0903 & 0.1397\\
    \multicolumn{2}{c}{\textbf{SISA}} & 0.0330 & 0.0389 & 0.0561 & 0.0795 & 0.0291 & 0.0346 & 0.0503 & 0.0723 & 0.0329 & 0.0414 & 0.0639 & 0.0974 \\
    \multicolumn{2}{c}{\textbf{SRU}} & \textbf{0.0370} & \textbf{0.0443} & \textbf{0.0642} & \textbf{0.0930} & \textbf{0.0348} & \textbf{0.0425} & \textbf{0.0632} & \textbf{0.0943} & \textbf{0.0419} & \textbf{0.0524} & \textbf{0.0802} & \textbf{0.1219} \\
    
    \bottomrule
    \end{tabular}
\end{table*}

\begin{table*}[t]
\caption{Unlearning effectiveness comparison. Lower scores denote better results. The best results are highlighted in bold.}
\label{table:effectiveness}
\centering
    \begin{tabular}{l cccc | cccc | cccc}
        \toprule
        \multicolumn{1}{l}{\multirow{2}*{\textbf{Beauty}}}  & \multicolumn{4}{c}{\textbf{GRU4Rec}} & \multicolumn{4}{c}{\textbf{SASRec}} & \multicolumn{4}{c}{\textbf{BERT4Rec}} \\
        \cmidrule(lr){2-5}\cmidrule(lr){6-9}\cmidrule(lr){10-13}
        \multicolumn{1}{l}{} & HIT@1 & HIT@5 & HIT@10 & HIT@20 & HIT@1 & HIT@5 & HIT@10 & HIT@20 & HIT@1 & HIT@5 & HIT@10 & HIT@20 \\
        \midrule
        \textbf{Retrain} & 0.0764 & 0.1715 & 0.2294 & 0.3052 & 0.0619 & 0.1566 & 0.2123 & 0.2807 & 0.0700 & 0.1588 & 0.2080 & 0.2739\\
        \textbf{SISA} & 0.0685 & 0.1654 & 0.2244 & 0.3074 & 0.0681 & 0.1605 & 0.2222 & 0.3091 & 0.0763 & 0.1730 & 0.2321 & 0.3119\\
        \textbf{SRU-R}& 0.0675 & 0.1561 & 0.2122 & 0.2809 & 0.0625 & 0.1468 & 0.2042 & 0.2697 & 0.0720 & 0.1573 & 0.2131 & 0.2798\\
        \textbf{SRU-C}& \textbf{0.0577} & \textbf{0.1335} & \textbf{0.1824} & \textbf{0.2510} & \textbf{0.0593} & \textbf{0.1429} & \textbf{0.1970} & \textbf{0.2666} & 0.0661 & \textbf{0.1516} & 0.2058 & \textbf{0.2689}\\
        \textbf{SRU-N}& 0.0643 & 0.1533 & 0.2028 & 0.2731 & 0.0605 & 0.1482 & 0.2039 & 0.2736 & \textbf{0.0638} & 0.1527 & \textbf{0.2054} & 0.2759\\
        \Xcline{1-1}{0.4pt}
        \Xhline{0.6pt}    

        \multicolumn{1}{l}{\multirow{2}*{\textbf{Steam}}}  & \multicolumn{4}{c}{\textbf{GRU4Rec}} & \multicolumn{4}{c}{\textbf{SASRec}} & \multicolumn{4}{c}{\textbf{BERT4Rec}} \\
        \cmidrule(lr){2-5}\cmidrule(lr){6-9}\cmidrule(lr){10-13}
        \multicolumn{1}{l}{} & HIT@1 & HIT@5 & HIT@10 & HIT@20 & HIT@1 & HIT@5 & HIT@10 & HIT@20 & HIT@1 & HIT@5 & HIT@10 & HIT@20 \\
        \midrule
        \textbf{Retrain}& 0.1581 & 0.3992 & 0.5372 & 0.6805 & 0.1411 & 0.3636 & 0.4975 & 0.6483 & 0.1159 & 0.3292 & 0.4701 & 0.6309\\
        \textbf{SISA}& 0.1582 & 0.3979 & 0.5349 & 0.6775 & 0.1410 & 0.3646 & 0.4959 & 0.6365 & 0.1166 & 0.3282 & 0.4668 & 0.6184\\
        \textbf{SRU-R}& 0.1545 & 0.3954 & 0.5319 & 0.6739 & 0.1412 & 0.3687 & 0.5020 & 0.6417 & 0.0992 & 0.2979 & 0.4282 & 0.5749\\
        \textbf{SRU-C}& 0.1499 & 0.3882 & 0.5241 & 0.6702 & 0.1389 & 0.3686 & 0.5041 & 0.6475 & 0.1036 & 0.3088 & 0.4407 & 0.5901\\
        \textbf{SRU-N}& \textbf{0.1461} & \textbf{0.3799} & \textbf{0.5136} & \textbf{0.6568} & \textbf{0.1138} & \textbf{0.3186} & \textbf{0.4422} & \textbf{0.5812} & \textbf{0.0957} & \textbf{0.2897} & \textbf{0.4205} & \textbf{0.5713}\\


        \bottomrule
    \end{tabular}
\end{table*}

\subsection{Unlearning Effectiveness (RQ2)}
In this part, we conduct experiments to evaluate the unlearning effectiveness of different methods.  
We randomly unlearn 10\% of data and set the extra deletion number $\mathcal{N}$ from 1 to 5. Table \ref{table:effectiveness} shows the unlearning effectiveness comparison on Beauty and Steam datasets. The results in the Games dataset show a similar conclusion. 

Firstly, we can see that even if the unlearning item is removed, there is a probability (e.g., more than 10\% on the Steam dataset) that the item can be inferred again from the remaining interactions in the session. This observation verifies that conventional exact unlearning methods cannot achieve exact unlearning effects in the session-based recommendation scenario.

Besides, we can see that the proposed SRU-R, SRU-C and SRU-N achieve better unlearning effectiveness compared with Retrain and SISA. For example, on the GRU4Rec model and trained on the Beauty dataset, the HIT@1 for SRU is 0.0577, while for Retrain is 0.0764. The observation indicates that the proposed data deletion module is essential for unlearning effectiveness. 

What's more, SRU-C and SRU-N achieve stable unlearning effectiveness improvement since they can help to eliminate the effect of collaborative correlations and sequential connections correspondingly, while SRU-R removes extra data randomly and has a more varied performance. 


To conclude, the proposed SRU achieves the highest unlearning effectiveness, even better than Retrain. 

\subsection{Unlearning Efficiency (RQ3)}
\begin{table*}
\caption{Comparison of unlearning efficiency (minute [m]). The best results are highlighted in bold.}
\label{table:efficency}
\centering
    \begin{tabular}{ll |ccc| ccc| ccc}
        \toprule
        \multicolumn{1}{l}{\textbf{Dataset}} & \multirow{2}*{} & \multicolumn{3}{c}{\textbf{Beauty}} & \multicolumn{3}{c}{\textbf{Games}} & \multicolumn{3}{c}{\textbf{Steam}} \\
        \midrule
        \textbf{Method} & & GRU4Rec & SASRec & BERT4Rec & GRU4Rec & SASRec & BERT4Rec & GRU4Rec & SASRec & BERT4Rec\\
        \midrule
        \multicolumn{1}{l}{\textbf{Retrain}}              &    & 46.80 & 55.6 & 55.76 & 31.22 & 29.91 & 31.14 & 274.67 & 368.99 & 296.89    \\
        \midrule
        \multicolumn{1}{l}{\multirow{3}{*}{\textbf{SRU}}} & \textbf{Sub-model} & 5.80 & 5.07 & 7.44 & 3.76 & 4.75 & 4.80 & 33.67 & 36.78 &  34.07     \\
                                                          & \textbf{Aggregation} & 0.72 & 6.05 & 5.53 & 1.78 & 4.40 & 3.87 & 25.30 & 62.53 &  64.30    \\
                                                          & \textbf{Total}& \textbf{6.52} & \textbf{11.12} & \textbf{12.97} & \textbf{5.54} & \textbf{9.15} & \textbf{8.67} & \textbf{58.97} & \textbf{99.31} &  \textbf{98.37}   \\
        \bottomrule
    \end{tabular}
    \vspace{-0.2cm}
\end{table*}

Table \ref{table:efficency} shows the training time comparison between Retrain and SRU. We evaluate them both on NVIDIA GeForce RTX 2080 Ti and set the shard number to 8. Especially the retraining time of SRU consists of sub-model training and aggregation module training.
From Table \ref{table:efficency}, we can find that SRU performs much more efficiently than Retrain. In most cases, SRU is more than three times faster than Retrain. For example, on Beauty and BERT4Rec, Retrain needs 55.76 minutes, but our SRU only needs 12.97 minutes. The efficiency improvement is more significant on the larger Steam dataset. For example, on Steam and SASRec, Retrain needs 368.99 minutes, but our SRU only needs 99.31 minutes, the improvement reaches 3.71x optimisation.
And the training of sub-models can be parallel since they do not share parameters which can further accelerate the training process.

\section{Conclusion}
In this paper, we have proposed a model-agnostic unlearning framework SRU for session-based recommendation. 
For an item-level unlearning request, SRU utilizes three data deletion strategies, including collaborative extra deletion (CED), neighbor extra deletion (NED), and random extra deletion (RED), to ensure the unlearned items cannot be inferred again from the remaining items in the session. 
Then we have retrained corresponding sub-model and the aggregation module for efficient unlearning. We have utilized a similarity-based session partition module and an attentive aggregation module to improve the recommendation performance in SRU. Besides, we have further defined an evaluation metric to verify the unlearning effectiveness of the session-based recommendation. 
We have implemented SRU with three representative session-based recommendation models and conducted experiments on three benchmark datasets from the perspective of unlearning performance, efficiency and effectiveness.
Experimental results have demonstrated the superiority of our proposed methods. 

For future work, we plan to investigate session-level unlearning. In the real-world recommendation scenario, both session-level and item-level unlearning need to be taken into consideration, and they may face different challenges. Furthermore, we would like to extend the exact deletion strategies to make more accuracy performance and completely unlearning. We also plan to adapt the approach to other recommendation domains. What's more, the trade-off between unlearning effectiveness, recommendation performance, and unlearning efficiency is also an interesting future topic. 
\section*{acknowledgement}
This work was supported by the Natural Science Foundation of China (62202271, 61902219, 61972234, 61672324, 62072279, 62102234, 62272274), the National Key R\&D Program of China with grants No. 2020YFB1406704 and No. 2022YFC3303004, the Natural Science Foundation of Shandong Province (ZR2021QF129, ZR2022QF004), the Key Scientific and Technological Innovation Program of Shandong Province (2019JZZY010129), the Fundamental Research Funds of Shandong University, the Tencent WeChat Rhino-Bird Focused Research Program (WXG-FR-2023-07). All content represents the opinion of the authors, which is not necessarily shared or endorsed by their respective employers and/or sponsors. This paper is dedicated to Mrs. Hui Yu. Thanks for her support to Xin Xin.

\clearpage



\clearpage
\bibliographystyle{ACM-Reference-Format}
\balance
\bibliography{references}
\clearpage
\appendix
\begin{figure*}[ht]
 \centering
 \vspace{0.1cm}
 \includegraphics[width=0.85\textwidth]{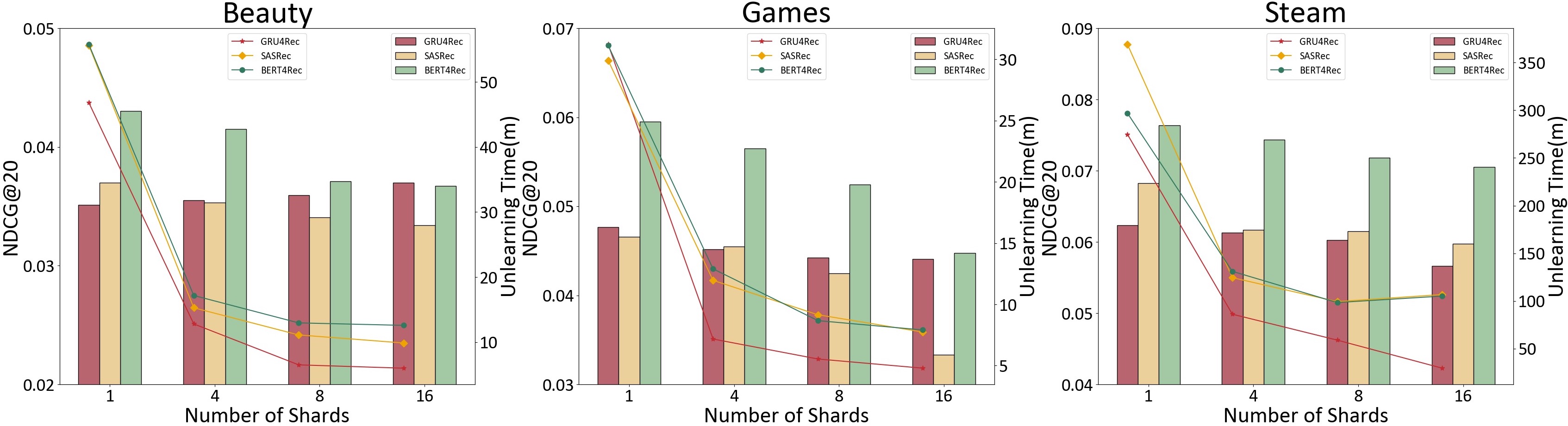}

 \caption{Impact of the shard number on recommendation performance and unlearning efficiency. The bar shows recommendation performance and the line shows unlearning time cost.}
\label{figure:shardnum}
\end{figure*}

\section{APPENDIX}
\subsection{Ablation Study}
\subsubsection{Shard numbers}

In this part, we conduct experiments to investigate the effect of shard numbers. Figure \ref{figure:shardnum} shows the results of NDCG@20 and the unlearning time cost with different shard numbers. We can see that a larger shard number leads to decreased recommendation performance and lower unlearning time costs. Since the sub-models are trained separately, a larger shard number indicates fewer correlations across sessions can be learned, resulting in lower recommendation performance. To this end, a good aggregation layer is needed to integrate the information of sub-models. The trade-off between recommendation performance and the unlearning cost is an interesting research direction.


\subsubsection{Session partition}

\begin{figure}[hb]
 \centering
 \vspace{0.1cm}
 \includegraphics[width=0.48\textwidth]{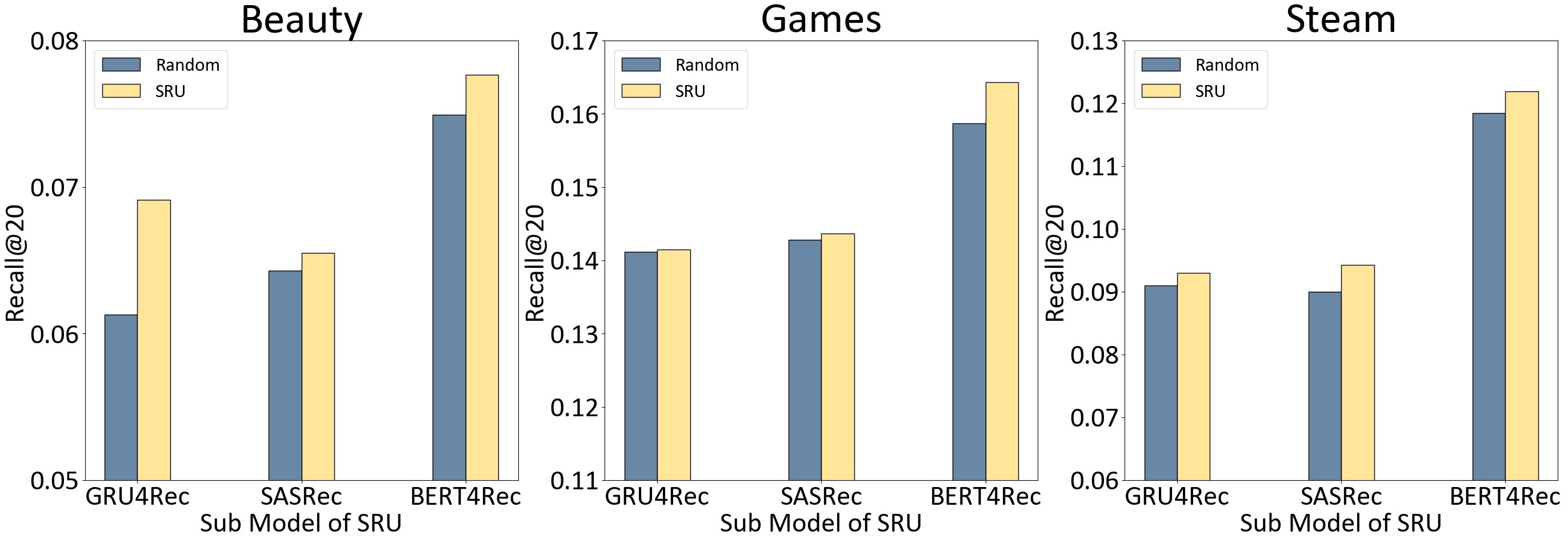}
 \caption{Effect of data partition.}
  \label{figure:random&k-means}
\end{figure}

In this part, we conduct experiments to verify the effect of the session partition module. 
Figure \ref{figure:random&k-means} 
shows the recommendation performance w.r.t. Recall@20 of the proposed partition method and the random partition method on the three datasets. We can observe that the proposed method achieves a much better recommendation performance compared to the random partition method. For instance, the Recall@20 of SRU is 0.069 on the Beauty dataset with the GRU4Rec model, while the corresponding Recall@20 of SISA is 0.061. This observation demonstrates the proposed session partition module helps to improve the recommendation performance.

\subsubsection{Data deletion}

\begin{figure}[hb]
 \centering
 \vspace{0.1cm}
 \includegraphics[width=0.48\textwidth]{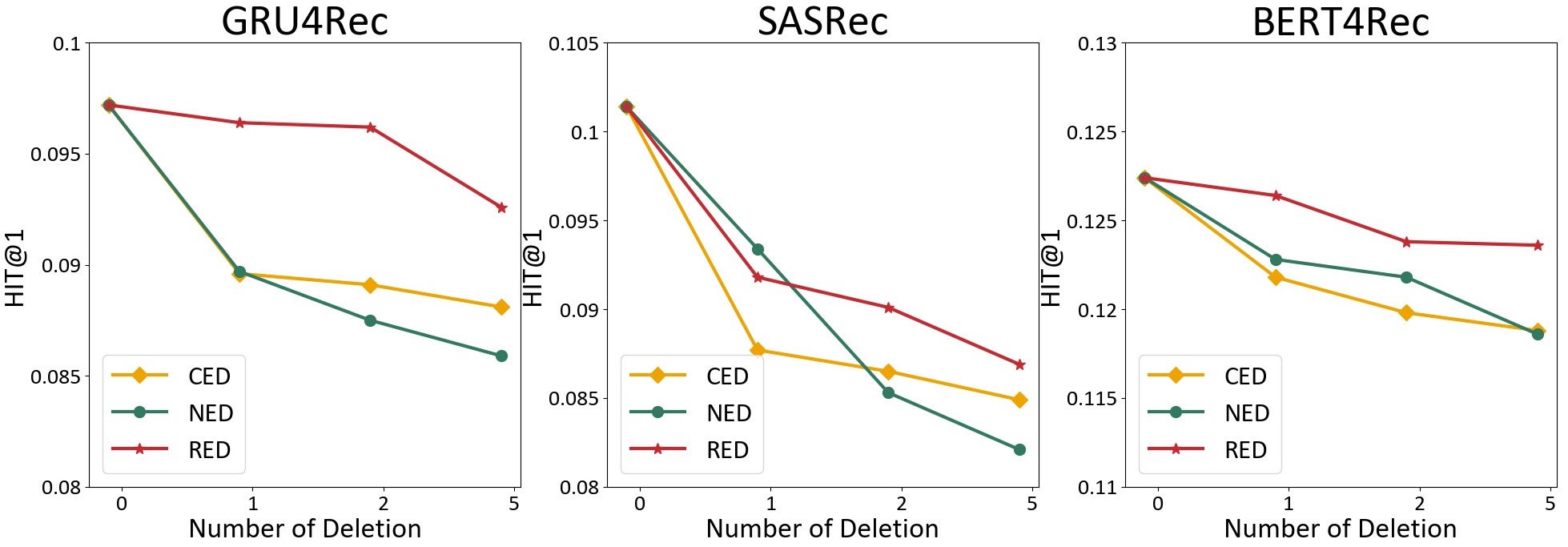}
 \caption{Effect of extra deletion numbers.}
 \label{figure:deletenum}
 \vspace{0.1cm}
\end{figure}

In this part, we conduct experiments to investigate how the number of extra deleted samples affects the unlearning effectiveness. 
We range the extra deletion number $\mathcal{N}$ from 0 to 5. 
Figure \ref{figure:deletenum} illustrates the results on the Games dataset. Results of other datasets show a similar trend. 
We can see that with the increase of the extra deletion number, the probability of inferring the unlearned items from the remaining data decreases. 
Intuitively, the increased extra deletion number could also downgrade the recommendation performance. In real-world applications, the trade-off between unlearning effectiveness and recommendation performance needs to be investigated more deeply.

\end{sloppypar}
\end{document}